%% file: main.tex
\pdfoutput=1
\documentclass[sigconf]{acmart}

\usepackage{subcaption}
\usepackage[subtle]{savetrees}
\usepackage{multirow}
\usepackage{microtype}
\usepackage{multicol}
\usepackage{pgfplots}
\usepackage{tabularx}
\usepackage{graphics}
\usepackage{adjustbox}
\usepackage{nicefrac}
\usepackage{booktabs}
\pgfplotsset{compat=1.11}
\usepgfplotslibrary{ternary}
\pgfplotsset{compat=newest}

\usepackage{setspace}
\usepackage{xspace} % insert space when needed.

\newcommand{\Modelone}{Score model\xspace}
\newcommand{\Modeltwo}{Rank model\xspace}
\newcommand{\Modelthree}{Rank\-Prob model\xspace}

\newcommand{\modelone}{\textit{score} model\xspace}
\newcommand{\modeltwo}{\textit{rank} model\xspace}
\newcommand{\modelthree}{\textit{rank\-prob} model\xspace}

\newcommand{\mone}{Score\xspace}
\newcommand{\mtwo}{Rank\xspace}
\newcommand{\mthree}{RankProb\xspace}

\newcommand{\Feedone}{Dense vector representation\xspace}
\newcommand{\Feedtwo}{Sparse vector representation\xspace}
\newcommand{\Feedthree}{Embedding vector representation\xspace}

\newcommand{\feedone}{dense vector representation\xspace}
\newcommand{\feedtwo}{sparse vector representation\xspace}
\newcommand{\feedthree}{embedding vector representation\xspace}

\newcommand{\fone}{Dense\xspace}
\newcommand{\ftwo}{Sparse\xspace}
\newcommand{\fthree}{Embed\xspace}

\newcommand{\shrink}{\vspace{-1.5ex}}
\newcommand{\sshrink}{\vspace{-.80ex}}

\def\:{\hskip0pt} % \: for hyphenation xxx\:---\:xxx 
\newcommand{\mypar}[1]{\vspace*{-0.1ex}\medskip\noindent\textbf{#1}~}
\newcounter{todocnt}

\definecolor{perp}{HTML}{A411CC}

\definecolor{cyan}{HTML}{307D7E}

\DeclareFontFamily{U}{mathb}{\hyphenchar\font45}
\DeclareFontShape{U}{mathb}{m}{n}{
<-6> mathb5 <6-7> mathb6 <7-8> mathb7
<8-9> mathb8 <9-10> mathb9
<10-12> mathb10 <12-> mathb12
}{}
\DeclareSymbolFont{mathb}{U}{mathb}{m}{n}

\DeclareMathSymbol{\smalltriangleup} {2}{mathb}{"98}% name to be checked
\DeclareMathSymbol{\smalltriangledown} {2}{mathb}{"99}% name to be checked
\DeclareMathSymbol{\smalltriangleleft} {2}{mathb}{"9A}% name to be checked
\DeclareMathSymbol{\smalltriangleright}{2}{mathb}{"9B}% name to be checked
\DeclareMathSymbol{\blacktriangleup} {2}{mathb}{"9C}% name to be checked
\DeclareMathSymbol{\blacktriangledown} {2}{mathb}{"9D}% name to be checked
\DeclareMathSymbol{\blacktriangleleft} {2}{mathb}{"9E}% name to be checked
\DeclareMathSymbol{\blacktriangleright}{2}{mathb}{"9F}% name to be checked
\copyrightyear{2017}
\acmYear{2017}
\setcopyright{rightsretained}
\acmConference{SIGIR '17}{August 07-11, 2017}{Shinjuku, Tokyo,
Japan}\acmDOI{10.1145/3077136.3080832}
\acmISBN{978-1-4503-5022-8/17/08}
\begin{document}
\title{Neural Ranking Models with Weak Supervision}

\author{Mostafa Dehghani}
\authornote{Work done while interning at Google Research.}
\affiliation{%
  %\institution{Institute for Logic, Language and Computation, University of Amsterdam}
  \institution{University of Amsterdam}
  %\streetaddress{}
  %\city{} 
  %\country{The Netherlands}
}
\email{dehghani@uva.nl}

\author{Hamed Zamani}
\affiliation{%
  %\institution{Center for Intelligent Information Retrieval, College of Information and Computer Sciences, University of Massachusetts Amherst}
 \institution{University of Massachusetts Amherst}
  %\streetaddress{}
  %\city{} 
  %\state{} 
  %\postcode{MA 01003}
  }
\email{zamani@cs.umass.edu}

\author{Aliaksei Severyn}
\affiliation{%
  \institution{Google Research}
  %\streetaddress{}
  %\city{} 
  %\country{The Netherlands}
}
\email{severyn@google.com}

\author{Jaap Kamps}
\affiliation{%
  \institution{University of Amsterdam}
  %\streetaddress{}
  %\city{} 
  %\country{The Netherlands}
}
\email{kamps@uva.nl}

\author{W. Bruce Croft}
\affiliation{%
  \institution{University of Massachusetts Amherst}
  %\streetaddress{}
  %\city{} 
  %\state{} 
  %\postcode{MA 01003}
  }
\email{croft@cs.umass.edu}

\renewcommand{\shortauthors}{M. Dehghani et al.}

\newcommand{\maingoal}{to study the impact of weak supervision on neural ranking models}
\newcommand{\rqone}{Is it possible to learn a neural ranker only from labels provided by a completely unsupervised IR model such as BM25, as the weak supervision signal, that will exhibit superior generalization capabilities?}
\newcommand{\rqtwo}{What input representation and learning objective is most suitable for learning in such a setting?}
\newcommand{\rqthree}{Can a supervised learning model benefit from a weak supervision step, especially in cases when labeled data is limited?}
\renewcommand{\rqone}{Can labels from an unsupervised IR model such as BM25 be used as weak supervision signal to train an effective neural ranker?}

\begin{abstract}
Despite the impressive improvements achieved by \emph{unsupervised} deep neural networks in computer vision and NLP tasks, such improvements have not yet been observed in ranking for information retrieval. The reason may be the complexity of the ranking problem, as it is not obvious how to learn from queries and documents when no supervised signal is available. Hence, in this paper, we propose to train a neural ranking model using \emph{weak supervision}, where labels are obtained automatically without human annotators or any external resources (e.g., click data). 
To this aim, we use the output of an unsupervised ranking model, such as BM25, as a weak supervision signal. We further train a set of simple yet effective ranking models based on feed-forward neural networks. We study their effectiveness under various learning scenarios (point-wise and pair-wise models) and using different input representations (i.e., from encoding query-document pairs into dense/sparse vectors to using word embedding representation). 
We train our networks using tens of millions of training instances and evaluate it on two standard collections: a homogeneous news collection (Robust)  and a heterogeneous large-scale web collection (ClueWeb). Our experiments indicate that employing proper objective functions and letting the networks to learn the input representation based on weakly supervised data leads to impressive performance, with over 13\% and 35\% MAP improvements over the BM25 model on the Robust and the ClueWeb collections. 
Our findings also suggest that supervised neural ranking models can greatly benefit from pre-training on large amounts of weakly labeled data that can be easily obtained from unsupervised IR models.

%\small
\mypar{KEYWORDS} ~~~ Ranking model, weak supervision, deep neural network, deep learning, ad-hoc retrieval
\end{abstract}

%\keywords{Ranking model, weak supervision, deep neural network, deep learning, ad-hoc retrieval


% We no longer use \terms command

%\keywords{Ranking model, weak supervision, deep neural network, deep learning, ad-hoc retrieval}

\maketitle
% \vfill
% \pagebreak

%\shrink\sshrink
\section{Introduction}
%General problem of need for massive data and pointing out the solution which is unsupervised learning:
Learning state-of-the-art deep neural network models requires a large amounts of labeled data, which is not always readily available and can be expensive to obtain. To circumvent the lack of human-labeled training examples, unsupervised learning methods aim to model the underlying data distribution, thus learning powerful feature representations of the input data, which can be helpful for building more accurate discriminative models especially when little or even no supervised data is available.

%On the success of the idea of unsupervised learning in other fields like NLP, or Vision
A large group of unsupervised neural models seeks to exploit the implicit internal structure of the input data, which in turn requires customized formulation of the training objective (loss function), targeted network architectures and often non-trivial training setups. 
% Not sure if we need all these examples, but they are nice to be included
For example in NLP, various methods for learning distributed word representations, e.g., word2vec~\citep{Mikolov:2013}, GloVe~\citep{Pennington:2014}, and sentence representations, e.g., paragraph vectors~\citep{Le:2014} and skip-thought~\citep{Kiros:2015} have been shown very useful to pre-train word embeddings that are then used for other tasks such as sentence classification, sentiment analysis, etc.
% Some examples to show that the new trend is completely toward unsupervised learning:
Other generative approaches such as language modeling in NLP, and, more recently, various flavors of auto-encoders~\citep{Baldi:2012} and generative adversarial networks~\citep{Goodfellow:2014} in computer vision have shown a promise in building more accurate models.

%Back to IR:
Despite the advances in computer vision, speech recognition, and NLP tasks using unsupervised deep neural networks, such advances have not been observed in core information retrieval (IR) problems, such as ranking. 
% Why it is not easy to do unsupervised learning for ranking problem:
A plausible explanation is the complexity of the ranking problem in IR, in the sense that it is not obvious how to learn a ranking model from queries and documents when no supervision in form of the relevance information is available.
% Ok.. the problem is clear. Considering this problem, what is the "general" idea of us in this paper:
To overcome this issue, in this paper, we propose to leverage large amounts of unsupervised data to infer ``noisy'' or ``weak'' labels and use that signal for learning supervised models as if we had the ground truth labels. 
% What we do in particular? -> using 
In particular, we use classic unsupervised IR models as a \emph{weak supervision} signal for training deep neural ranking models.
% Before going forward and talking about the idea itself, let's just explain the "weak supervision" in a simple way.
Weak supervision here refers to a learning approach that creates its own training data by heuristically retrieving documents for a large query set. 
This training data is created automatically, and thus it is possible to generate billions of training instances with almost no cost.\footnote{Although weak supervision may refer to using noisy data, in this paper, we assume that no external information, e.g., click-through data, is available.}
As training deep neural networks is an exceptionally data hungry process, the idea of pre-training on massive amount of weakly supervised data and then fine-tuning the model using a small amount of supervised data could improve the performance~\citep{Rrhan:2010}. 

% JK: General blob moved to the end of the section.

%The main research questions addressed in this paper are the following:
The main aim of this paper is \textsl{\maingoal}, which we break down into the following concrete research questions:
\begin{enumerate}
  \setlength{\topsep}{0pt}
  \setlength{\partopsep}{0pt}
  \setlength{\itemsep}{0pt}
  \setlength{\parskip}{0pt}
  \setlength{\parsep}{0pt}
\item[\textbf{RQ1}] \textsl{\rqone}
\item[\textbf{RQ2}] \textsl{\rqtwo}
\item[\textbf{RQ3}] \textsl{\rqthree}
\end{enumerate}
%\sshrink

% short reports on the experimental results and findings
We examine various neural ranking models with different ranking architectures and objectives, i.e., point-wise and pair-wise, as well as different input representations, from encoding query-document pairs into dense\:/\:sparse vectors to learning query\:/\:document embedding representations. 
The models are trained on billions of training examples that are annotated by BM25, as the weak supervision signal.
%Simply mentioning that we have successful result, just  py pointing out that we can go beyond performance of BM25
Interestingly, we observe that using just training data that are annotated by BM25 as the weak annotator, we can outperform BM25 itself on the test data.
Based on our analysis, the achieved performance is generally indebted to three main factors: 
First, defining an objective function that aims to learn the ranking instead of calibrated scoring to relax the network from fitting to the imperfections in the weakly supervised training data.
Second, letting the neural networks learn optimal query/document representations instead of feeding them with a representation based on predefined features. This is a key requirement to maximize the benefits from deep learning models with weak supervision as it enables them to generalize better.
%
%Third and last, training the network on a massive amount of training data, which is possible with weak supervision setting.
Third and last, the weak supervision setting makes it possible to train the network on a massive amount of training data.

We further thoroughly analyse the behavior of models to understand what they learn, what is the relationship among different models, and how much training data is needed to go beyond the weak supervision signal. We also study if employing deep neural networks may help in different situations. 
Finally, we examine the scenario of using the network trained on a weak supervision signal as a pre-training step. We demonstrate that, in the ranking problem, the performance of deep neural networks trained on a limited amount of supervised data significantly improves when they are initialized from a model pre-trained on weakly labeled data.

%what is cool about weak supervision in terms of using traditional IR models? 
Our results have broad impact as the proposal to use unsupervised traditional methods as weak supervision signals is applicable to variety of IR tasks, such as filtering or classification, without the need for supervised data. 
More generally, our approach unifies the classic IR models with currently emerging data-driven approaches in an elegant way.

\shrink
\section{Related Work}
Deep neural networks have shown impressive performance in many computer vision, natural language processing, and speech recognition tasks~\citep{Lecun:2015}. 
Recently, several attempts have been made to study deep neural networks in IR applications, which can be generally partitioned into two categories~\citep{Onal:2016, Zhang:2016}. 
The first category includes approaches that use the results of trained (deep) neural networks in order to improve the performance in IR applications. Among these, distributed word representations or embeddings~\citep{Mikolov:2013,Pennington:2014} have attracted a lot of attention. Word embedding vectors have been applied to term re-weighting in IR models~\citep{Zheng:2015,Rekabsaz:2017}, query expansion~\citep{Diaz:2016,Zamani:2016a,Rekabsaz:2016}, query classification~\citep{Liu:2015,Zamani:2016b},  etc. 
The main shortcoming of most of the approaches in this category is that the objective of the trained neural network differs from the objective of these tasks.  For instance, the word embedding vectors proposed in~\citep{Mikolov:2013,Pennington:2014} are trained based on term proximity in a large corpus, which is different from the objective in most IR tasks. 
To overcome this issue, some approaches try to learn representations in an end-to-end neural model for learning a specific task like entity ranking for expert finding~\citep{VanGysel:2016:www} or product search~\citep{VanGysel:2016:cikm}. \citet{Zamani:2017} recently proposed relevance-based word embedding models for learning word representations based on the objectives that matter for IR applications.

The second category, which this paper belongs to, consists of the approaches that design and train a (deep) neural network for a specific task, e.g., question answering~\citep{Cohen:2016,Yang:2016}, click models~\citep{Borisov:2016}, context-aware ranking~\citep{Zamani:2017b}, etc.
A number of the approaches in this category have been proposed for ranking documents in response to a given query.
These approaches can be generally divided into two groups: \emph{late combination models} and \emph{early combination models} (or representation-focused and interaction-focused models according to~\citep{Guo:2016}). 
The late combination models, following the idea of Siamese networks~\citep{Bromley:1993}, independently learn a representation for each query and candidate document and then calculate the similarity between the two estimated representations via a similarity function. For example, \citet{Huang:2013} proposed DSSM, which is a feed forward neural network with a word hashing phase as the first layer to predict the click probability given a query string and a document title. 
The DSSM model was further improved by incorporating convolutional neural networks~\citep{Shen:2014}.

On the other hand, the early combination models are designed based on the interactions between the query and the candidate document as the input of network. 
For instance, DeepMatch~\citep{Lu:2013} maps each text to a sequence of terms and trains a feed-forward network for computing the matching score. 
The deep relevance matching model for ad-hoc retrieval~\citep{Guo:2016} is another example of an early combination model that feeds a neural network with the  histogram-based features representing interactions between the query and document. 
Early combining enables the model to have an opportunity to capture various interactions between query and document(s), while with late combination approach, the model has only the chance of isolated observation of input elements. Recently, Mitra et al.~\citep{Mitra:2016} proposed to simultaneously learn local and distributional representations, which are early and late combination models respectively,  to capture both exact term matching and semantic term matching.

Until now, all the proposed neural models for ranking are trained on either explicit relevance judgements or clickthrough logs. However, a massive amount of such training data is not always available. 

In this paper, we propose to train neural ranking models using weak supervision, which is the most natural way to reuse the existing supervised learning models where the imperfect labels are treated as the ground truth.
The basic assumption is that we can cheaply obtain labels (that are of lower quality than human-provided labels) by expressing the prior knowledge we have about the task at hand by specifying a set of heuristics, adapting existing ground truth data for a different but related task (this is often referred to distant supervision\footnote{We do not distinguish between weak and distant supervision as the difference is subtle and both terms are often used interchangeably in the literature.}), extracting supervision signal from external knowledge-bases or ontologies, crowd-sourcing partial annotations that are cheaper to get, etc.
Weak supervision is a natural way to benefit from unsupervised data and it has been applied in NLP for various tasks including relation extraction~\citep{Bing:2015,Han:2016}, knowledge-base completion~\citep{Hoffmann:2011}, sentiment analysis~\citep{Severyn:2015}, etc.  
There are also similar attempts in IR for automatically constructing test collections~\citep{Asadi:2011} and learning to rank using labeled features, i.e. features that an expert believes they are correlated with relevance~\citep{Diaz:2016:ictir}.
In this paper, we make use of traditional IR models as the weak supervision signal to generate a large amount of training data and train effective neural ranking models that outperform the baseline methods by a significant margin.

\sshrink
\section{Weak Supervision for Ranking}
Deep learning techniques have taken off in many fields, as they automate the onerous task of input representation and feature engineering. 
On the other hand, the more the neural networks become deep and complex, the more it is crucial for them to be trained on massive amounts of training data.
In many applications, rich annotations are costly to obtain and task-specific training data is now a critical bottleneck. 
Hence, unsupervised learning is considered as a long standing goal for several applications. However, in a number of information retrieval tasks, such as ranking, it is not obvious how to train a model with large numbers of queries and documents with no relevance signal.
To address this problem in an unsupervised fashion, we use the idea of ``Pseudo-Labeling'' by taking advantage of existing unsupervised methods for creating a weakly annotated set of training data and we propose to train a neural retrieval model with weak supervision signals we have generated.
In general, weak supervision refers to learning from training data in which the labels are imprecise. In this paper, we refer to weak supervision as a learning approach that automatically creates its own training data using an existing unsupervised approach, which differs from imprecise data coming from external observations (e.g., click-through data) or noisy human-labeled data.

We focus on query-dependent ranking as a core IR task. To this aim, we take a well-performing existing unsupervised retrieval model, such as BM25. This model plays the role of ``pseudo-labeler'' in our learning scenario. In more detail, given a target collection and a large set of training queries (without relevance judgments), we make use of the pseudo-labeler to rank/score the documents for each query in the training query set.  Note that we can generate as much as training data as we need with almost no cost. The goal is to train a ranking model given the scores/ranking generated by the pseudo-labeler as a weak supervision signal. 
 
In the following section, we formally present a set of neural network-based ranking models that can leverage the given weak supervision signal in order to learn accurate representations and ranking for the ad-hoc retrieval task.

\sshrink
\section{Neural Ranking Models}
In this section, we first introduce our ranking models. Then, we describe the architecture of the base neural network model shared by different ranking models. Finally, we discuss the three input layer architectures used in our neural rankers to encode (query, candidate document) pairs.

\label{sec:models}
\begin{figure}[t]
    \centering
    \begin{subfigure}[t]{0.23\columnwidth}
        \centering
        \includegraphics[height=3.2cm]{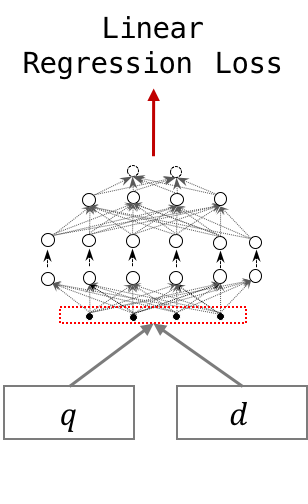}\vspace*{-3ex}%
        \caption{\label{fig:m1}\mone model}
    \end{subfigure}%
    ~
    \begin{subfigure}[t]{0.40\columnwidth}
        \centering
        \includegraphics[height=3.2cm]{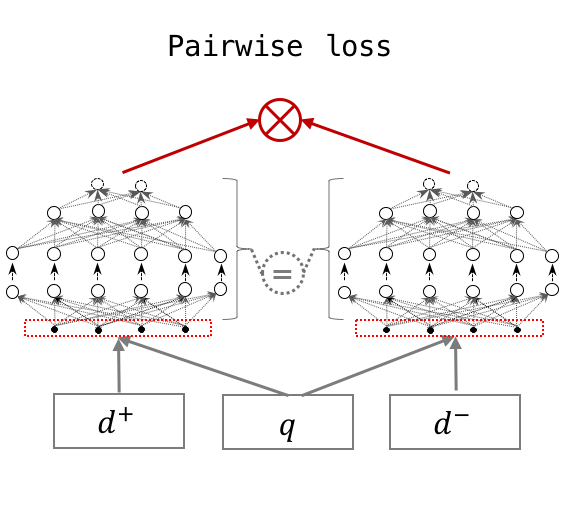}\vspace*{-3ex}%
        \caption{\label{fig:m2}\mtwo model}
    \end{subfigure}%
    ~
    \begin{subfigure}[t]{0.37\columnwidth}
        \centering
        \includegraphics[height=3.2cm]{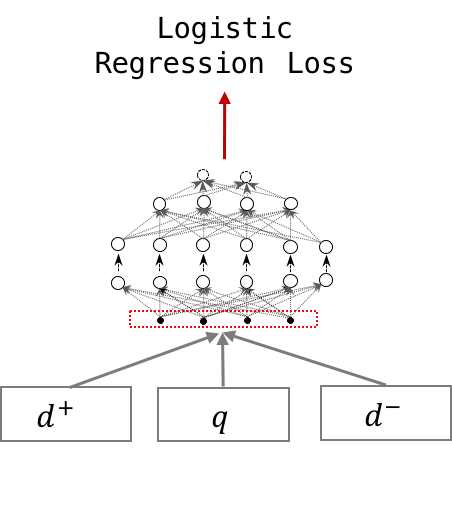}\vspace*{-3ex}%
        \caption{\label{fig:m3}\mthree model}
    \end{subfigure}%
    \vspace*{-2ex}%
    \caption{\label{fig:ranking-arch} Different Ranking Architectures}
    \vspace{-20pt}
\end{figure}
\sshrink
\subsection{Ranking Architectures}
We define three different ranking models: 
one point-wise and two pair-wise models. 
We introduce the architecture of these models and explain how we train them using weak supervision signals.

\mypar{\Modelone}: This architecture models a point-wise ranking model that learns to predict retrieval scores for query-document pairs. More formally, the goal in this architecture is to learn a \emph{scoring function} $\mathcal{S}(q, d; \theta)$ that determines the retrieval score of document $d$ for query $q$, given a set of model parameters $\theta$.
In the training stage, we are given a training set comprising of training instances each a triple $\tau = (q,d, s_{q,d})$, where $q$ is a query from training query set $Q$, $d$ represents a retrieved document for the query $q$, and $s_{q,d}$ is the relevance score (calculated by a weak supervisor), which is acquired using a retrieval scoring function in our setup.
We consider the mean squared error as the loss function for a given batch of training instances:
\begin{equation}
\mathcal{L}(b; \theta) = \frac{1}{|b|} \sum_{i=1}^{|b|}{(\mathcal{S}(\{q, d\}_i; \theta) - s_{\{q, d\}_i})^2}
\end{equation}
where $\{q, d\}_i$ denotes the query and the corresponding retrieved document in the $i^{th}$ training instance, i.e. $\tau_i$ in the batch $b$.
The conceptual architecture of the model is illustrated in Figure~\ref{fig:m1}.

\mypar{\Modeltwo}:
In this model, similar to the previous one, the goal is to learn a scoring function $\mathcal{S}(q, d; \theta)$ for a given pair of query $q$ and document $d$ with the set of model parameters $\theta$. 
However, unlike the previous model, we do not aim to learn a calibrated scoring function. 
In this model, as it is depicted in Figure~\ref{fig:m2}, we use a pair-wise scenario during training in which we have two point-wise networks that share parameters and we update their parameters to minimize a pair-wise loss.
In this model, each training instance has five elements: $\tau = (q,d_1, d_2, s_{q,d_1}, s_{q,d_2})$.
During the inference, we treat the trained model as a point-wise scoring function to score query-document pairs.

We have tried different pair-wise loss functions and empirically found that the model learned based on the hinge loss (max-margin loss function) performs better than the others. 
Hinge loss is a linear loss that penalizes examples that violate the margin constraint. It is widely used in various learning to rank algorithms, such as Ranking SVM~\citep{Herbrich:1999}. The hinge loss function for a batch of training instances is defined as follows:
\begin{equation}
\begin{aligned}
\mathcal{L}(b; \theta) = \frac{1}{|b|}
\sum_{i=1}^{|b|}
\max\big\{
& 
0, \varepsilon - \text{sign}
(s_{\{q, d_1\}_i} - s_{\{q, d_2\}_i})
& \\ & 
\left(\mathcal{S}\left(\{q, d_1\}_i; \theta\right) -\mathcal{S}\left(\{q, d_2\}_i; \theta\right)\right)
\big\}
, 
%\nonumber
\end{aligned}     
\end{equation}
where $\varepsilon$ is the parameter determining the margin of hinge loss. We found that as we compress the outputs to the range of $[-1, 1]$, $\varepsilon=1$ works well as the margin for the hinge loss function.

\mypar{\Modelthree}:
The third architecture is based on a pair-wise scenario during both training and inference (Figure~\ref{fig:m3}). This model learns a \emph{ranking function} $\mathcal{R}(q, d_1, d_2; \theta)$ which predicts the probability of document $d_1$ to be ranked higher than $d_2$ given $q$.
Similar to the \modeltwo, each training instance has five elements: $\tau = (q,d_1, d_2, s_{q,d_1}, s_{q,d_1})$.
For a given batch of training instances, we define our loss function based on cross-entropy as follows:
\begin{align}
\mathcal{L}(b; \theta) = -\frac{1}{|b|}
\sum_{i=1}^{|b|} &
P_{\{q,d_1,d_2\}_i} \log(\mathcal{R}(\{q,d_1,d_2\}_i; \theta)) \\
&
+ (1- P_{\{q,d_1,d_2\}_i})\log(1- \mathcal{R}(\{q,d_1,d_2\}_i; \theta)) \nonumber
\end{align}
where $P_{\{q,d_1,d_2\}_i}$ is the probability of document $d_1$ being ranked higher than $d_2$, based on the scores obtained from training instance $\tau_i$:
\begin{equation}
P_{\{q,d_1,d_2\}_i} = \frac{s_{\{q,d_1\}_i}}{s_{\{q,d_1\}_i} + s_{\{q,d_2\}_i}}
\end{equation}

%A similar loss function has been previously used in RankNet~\citep{Burges:2005}.
It is notable that at inference time, we need a scalar score for each document. Therefore, we need to turn the model's pair-wise predictions into a score per document. To do so, for each document, we calculate the average of predictions against all other candidate documents, which has $O(n^2)$ time complexity and is not practical in real-world applications. There are some approximations could be applicable to decrease the time complexity at inference time~\citep{Wauthier:2013}.

\sshrink
\subsection{Neural Network Architecture}
As shown in Figure~\ref{fig:ranking-arch}, all the described ranking architectures share a neural network module. 
We opted for a simple feed-forward neural network which is composed of: input layer $z_0$, $l-1$ hidden layers, and the output layer $z_l$. The input layer $z_0$ provides a mapping $\psi$ to encode the input query and document(s) into a fixed-length vector.
The exact specification of the input representation feature function $\psi$ is given in the next subsection. 
Each hidden layer $z_i$ is a fully-connected layer that computes the following transformation:
\begin{equation}
    z_i = \alpha(W_i.z_{i-1} + b_i); ~ 1<i<l-1,
\end{equation}
where $W_i$ and $b_i$ respectively denote the weight matrix and the bias term corresponding to the $i^{th}$ hidden layer, and $\alpha(.)$ is the activation function. We use the rectifier linear unit $\textit{ReLU}(x) = \max(0, x)$ as the activation function, which is a common choice in the deep learning literature~\citep{Lecun:2015}. 
The output layer $z_l$ is a fully-connected layer with a single continuous output. The activation function for the output layer depends on the ranking architecture that we use. For the \modelone architecture, we empirically found that a linear activation function works best, while $tanh$ and the sigmoid functions are used for the \modeltwo and \modelthree respectively.

%To optimize our neural ranking models we use the SGD. 
Furthermore, to prevent feature co-adaptation, we use \emph{dropout}~\citep{Srivastava:2014} as the regularization technique in all the models. Dropout sets a portion of hidden units to zero during the forward phase when computing the activations which prevents overfitting.

%In this paper, we experiment with simple feed-forward neural nets. Using more expressive architectures such as CNNs or LSTMs is outside of the scope of this paper and is our future work.
%

\sshrink
\subsection{Input Representations}
\label{sec:feedings}
We explore three definitions of the input layer representation $z_0$ captured by a feature function $\psi$ that maps the input into a fixed-size vector which is further fed into the fully connected layers: 
(i) a conventional dense feature vector representation that contains various statistics describing the input query-document pair, 
(ii) a sparse vector containing bag-of-words representation, and 
(iii) bag-of-embeddings averaged with learned weights. 
These input representations define how much capacity is given to the network to extract discriminative signal from the training data and thus result in different generalization behavior of the networks. 
It is noteworthy that input representation of the networks in the \modelone and \modeltwo is defined for a pair of the query and the document, while the network in the \modelthree needs to be fed by a triple of the query, the first document, and the second document.

\mypar{\Feedone (\fone)}: 
In this setting, we build a dense feature vector composed of features used by traditional IR methods, e.g., BM25. The goal here is to let the network fit the function described by the BM25 formula when it receives exactly the same inputs. 
In more detail, our input vector is a concatenation ($||$) of the following inputs: total number of documents in the collection (i.e., $N$), average length of documents in the collection (i.e., $avg(l_d)_D$), document length (i.e., $l_d$), frequency of each query term $t_i$ in the document (i.e., $tf(t_i, d)$), and document frequency of each query term (i.e., $df(t_i)$). Therefore, for the point-wise setting, we have the following input vector:
\begin{equation}
\psi(q, d) = [N || avg(l_d)_D || l_d || \{df(t_i) || tf(t_i,d)\}_{1 \leq i \leq k}],
\end{equation}
where $k$ is set to a fixed value ($5$ in our experiments). 
We truncate longer queries and do zero padding for shorter queries. 
For the networks in the \modelthree, we consider a similar function with additional elements: the length of the second document and the frequency of query terms in the second document.
%I changed the description slightly and I guess there is no need to bring the equation again for pair-wise!
% \begin{equation}
% \psi(q, d) = [N || avg(l_d)_D || l_d_1 || \{df(t_i) || tf(t_i,d_1) || tf(t_i,d_2)\}_{1 \leq i \leq k}],
% \end{equation}

\mypar{\Feedtwo (\ftwo)}: 
Next, we move away from a fully featurized representation that contains only aggregated statistics and let the network performs feature extraction for us. In particular, we build a bag-of-words representation by extracting term frequency vectors of query ($tfv_q$), document ($tfv_d$), and the collection ($tfv_c$) and feed the network with concatenation of these three vectors. For the point-wise setting, we have the following input vector:
\begin{equation}
\psi(q, d) = [tfv_c || tfv_q || tfv_d]
\end{equation}
For the network in \modelthree, we have a similar input vector with both $tfv_{d_1}$ and $tfv_{d_2}$. Hence, the size of the input layer is $3 \times vocab~size$ in the point-wise setting, and $4 \times vocab~size$ in the pair-wise setting. 
%In this paradigm, we have no document-level statistic, e.g., document frequency, although collection term frequency would be considered as a clue for term weighting. 

\mypar{\label{sec:feedthree}\Feedthree (\fthree)}:
The major weakness of the previous input representation is that words are treated as discrete units, hence prohibiting the network from  performing soft matching between semantically similar words in queries and documents. In this input representation paradigm, we rely on word embeddings to obtain more powerful representation of queries and documents that could bridge the lexical chasm. % of \fthree.
The representation function $\psi$ consists of three components: an embedding function $\mathcal{E}: \mathcal{V} \rightarrow \mathbb{R}^{m}$ (where $\mathcal{V}$ denotes the vocabulary set and $m$ is the embedding dimension), a weighting function $\mathcal{W}: \mathcal{V} \rightarrow \mathbb{R}$, and a compositionality function $\odot: (\mathbb{R}^{m}, \mathbb{R})^n \rightarrow \mathbb{R}^{m}$. More formally, the function $\psi$ for the point-wise setting is defined as:
\begin{equation}
\psi(q, d) = [\odot_{i=1}^{|q|}(\mathcal{E}(t_i^q), \mathcal{W}(t_i^q)) || \odot_{i=1}^{|d|} (\mathcal{E}(t_i^d), \mathcal{W}(t_i^d))],
\end{equation}
where $t_i^q$ and $t_i^d$ denote the $i^{th}$ term in query $q$ and document $d$, respectively. 
For the network of the \modelthree, another similar term is concatenated with the above vector for the second document. The embedding function $\mathcal{E}$ transforms each term to a dense $m$-\:dimensional float vector as its representation, which is learned during the training phase. The weighting function $\mathcal{W}$ assigns a weight to each term in the vocabulary set, which is supposed to learn term global importance for the retrieval task. The compositionality function $\odot$ projects a set of $n$ embedding and weighting pairs to an $m$-\:dimensional representation, independent from the value of $n$. The compositionality function is given by:
\begin{equation}
\odot_{i=1}^n(\mathcal{E}(t_i), \mathcal{W}(t_i)) = \sum_{i=1}^n \widehat{\mathcal{W}}(t_i)\cdot \mathcal{E}(t_i),
\end{equation}
which is the weighted element-wise sum of the terms' embedding vectors. $\widehat{\mathcal{W}}$ is the normalized weight that is learned for each term, given as follows:
\begin{equation}
\widehat{\mathcal{W}}(t_i) = \frac{\exp(\mathcal{W}(t_i))}{\sum_{j=1}^n{ \exp(\mathcal{W}(t_j))}}
\end{equation}

\medskip
All combinations of different ranking architectures and different input representations presented in this section can be considered for developing ranking models.

\sshrink
\section{Experimental Design}
In this section, we describe the train and evaluation data, metrics we report, and detailed experimental setup. Then we discuss the results.

\sshrink
\subsection{Data}
\label{sec:data}
\mypar{Collections.}
In our experiments, we used two standard TREC collections: The first collection (called \emph{Robust04}) consists of over 500k news articles from different news agencies, that is available in TREC Disks 4 and 5 (excluding Congressional Records). This collection, which was used in TREC Robust Track 2004, is considered as a homogeneous collection, because of the nature and the quality of documents. The second collection (called \emph{ClueWeb}) that we used is ClueWeb09 Category B, a large-scale web collection with over 50 million English documents, which is considered as a heterogeneous collection. This collection has been used in TREC Web Track, for several years. In our experiments with this collection, we filtered out the spam documents using the Waterloo spam scorer\footnote{\url{http://plg.uwaterloo.ca/~gvcormac/clueweb09spam/}}~\citep{Cormack:2011} with the default threshold $70\%$. The statistics of these collections are reported in Table~\ref{tab:data}. 

\mypar{Training query set.}
To train our neural ranking models, we used the unique queries (only the query string) appearing in the AOL query logs~\citep{Pass:2006}. This query set contains web queries initiated by real users in the AOL search engine that were sampled from a three-month period from March 1, 2006 to May 31, 2006. We filtered out a large volume of navigational queries containing URL substrings (``http'', ``www.'', ``.com'', ``.net'', ``.org'', ``.edu''). We also removed all non-alphanumeric characters from the queries. We made sure that no queries from the training set appear in our evaluation sets. For each dataset, we took queries that have at least ten hits in the target corpus using the pseudo-labeler method. Applying all these processes, we ended up with 6.15 million queries for the Robust04 dataset and 6.87 million queries for the ClueWeb dataset. 
In our experiments, we randomly selected $80\%$ of the training queries as training set and the remaining $20\%$ of the queries were chosen as validation set for hyper-parameter tuning. As the ``pseudo-labeler'' in our training data, we have used BM25 to score/rank documents in the collections given the queries in the training query set.

\mypar{Evaluation query sets.} 
We use the following query sets for evaluation that contain human-labeled judgements: a set of 250 queries (TREC topics 301--450 and 601--700) for the Robust04 collection that were previously used in TREC Robust Track 2004. A set of 200 queries (topics 1-200) were used for the experiments on the ClueWeb collection. These queries were used in TREC Web Track 2009--2012. We only used the title of topics as queries.

\input{table_data}

\subsection{Evaluation Metrics.}
To evaluate retrieval effectiveness, we report three standard evaluation metrics: mean average precision (MAP) of the top-ranked $1000$ documents, precision of the top $20$ retrieved documents (P@20), and normalized discounted cumulative gain (nDCG)~\citep{Jarvelin:2002} calculated for the top $20$ retrieved documents (nDCG@20). Statistically significant differences of MAP, P@20, and nDCG@20 values are determined using the two-tailed paired t-test with $p\_value<0.05$, with Bonferroni correction.
\input{table_res} 

\sshrink
\subsection{Experimental Setup}
All models described in Section~\ref{sec:models} are implemented using TensorFlow~\citep{tang2016:tflearn,tensorflow2015-whitepaper}.
In all experiments, the parameters of the network are optimized employing the Adam optimizer~\citep{Kingma:2014} and using the computed gradient of the loss to perform the back-propagation algorithm.
All model hyper-parameters were tuned on the respective validation set (see Section~\ref{sec:data} for more detail) using batched GP bandits with an expected improvement acquisition function~\citep{Desautels:2014}. 
For each model, the size of hidden layers and the number of hidden layers were selected from $[16, 32, 64, 128, 256, 512, 1024]$ and $[1, 2, 3, 4]$, respectively. The initial learning rate and the dropout parameter were selected from $[1E-3, 5E-4, 1E-4, 5E-5, 1E-5]$ and $[0.0, 0.1, 0.2, 0.5]$, respectively. For models with \feedthree, we considered embedding sizes of $[100, 300, 500, 1000]$. As the training data, we take the top $1000$ retrieved documents for each query from training query set $Q$, to prepare the training data. In total, we have $|Q|\times 1000$ ($\sim6E10$ examples in our data) point-wise example and $\sim|Q|\times 1000^2$ ($\sim6E13$ examples in our data) pair-wise examples. The batch size in our experiments was selected from  $[128, 256, 512]$.
At inference time, for each query, we take the top $2000$ retrieved documents using BM25 as candidate documents and re-rank them by the trained models. In our experiments, we use the Indri\footnote{\url{https://www.lemurproject.org/indri.php}} implementation of BM25 with the default parameters (i.e., $k_1 = 1.2$, $b = 0.75$, and $k_3 = 1000$).

%We open source all the implementations of models and the codes for replicating training query set, which are available online\footnote{Link to the repository is removed to preserve anonymity.}.

\shrink
\section{Results and Discussion}
In the following, we evaluate our neural rankers trained with different learning approaches (Section~\ref{sec:models}) and different input representations (Section~\ref{sec:feedings}). We attempt to break down our research questions to several subquestions, and provide empirical answers along with the intuition and analysis behind each question:

\mypar{How do the neural models with different training objectives and input representations compare?}
Table~\ref{tbl_main} presents the performance of all model combinations.
Interestingly, combinations of the \modeltwo and the \modelthree with \feedthree outperform BM25 by significant margins in both collections. For instance, the \modelthree with \feedthree that shows the best performance among the other methods, surprisingly, improves BM25 by over $13\%$ and $35\%$ in Robust04 and ClueWeb collections respectively, in terms of MAP. Similar improvements can be observed for the other evaluation metrics.

Regarding the modeling architecture, in the \modeltwo and the \modelthree, compared to the \modelone, we define objective functions that target to learn ranking instead of scoring. This is particularly important in weak supervision, as the scores are imperfect values\:---\:using the ranking objective alleviates this issue by forcing the model to learn a preference function rather than reproduce absolute scores.
In other words, using the ranking objective instead of learning to predict calibrated scores allows the \modeltwo and the \modelthree to learn to distinguish between examples whose scores are close. This way, some small amount of noise, which is a common problem in weak supervision, would not perturb the ranking as easily.

Regarding the input representations, \feedthree leads to better performance compared to the other ones in all models.
Using \feedthree not only provides the network with more information, but also lets the network to learn proper representation capturing the needed elements for the next layers with better understanding of the interactions between query and documents. 
Providing the network with already engineered features would block it from going beyond the weak supervision signal and limit the ability of the models to learn latent features that are unattainable through feature engineering. 

Note that although the \modelthree is more precise in terms of MAP, the \modeltwo is much faster in the inference time ($O(n)$ compared to $O(n^2)$), which is a desirable property in real-life applications.

\input{plot_loss_step}
\input{models_proximity}

\mypar{Why do \feedone and \feedtwo fail to replicate the performance of BM25?}
Although neural networks are capable of approximating arbitrarily complex non-linear functions, we observe that the models with \feedone fail to replicate the BM25 performance, while they are given the same feature inputs as the BM25 components (e.g., TF, IDF, average document length, etc). To ensure that the training converges and there is no overfitting, we have looked into the training and validation loss values of different models during the training time. Figure~\ref{fig:step-loss} illustrates the loss curves for the training and validation sets (see Section~\ref{sec:data}) per training step for different models.
As shown, in models with \feedone, the training losses drop quickly to values close to zero while this is not the case for the validation losses, which is an indicator of over-fitting on the training data. 
Although we have tried different regularization techniques, like $l_2$-regularization and dropout with various parameters, there is less chance for generalization when the networks are fed with the fully featurized input. Note that over-fitting would lead to poor performance, especially in weak supervision scenarios as the network learns to model imperfections from weak annotations. 
This phenomenon is also the case for models with the \feedtwo, but with less impact. However, in the models with the \feedthree, the networks do not overfit, which helps it to go beyond the weak supervision signals in the training data.

\mypar{How are the models related?}
To better understand the relationship of different neural models described above, we compare their performance across the query dimension following the approach in~\citep{Mitra:2016}. 
We assume that similar models should perform similarly for the same queries. Hence, we represent each model by a vector, called the performance vector, whose elements correspond to per query performance of the model, in terms of nDCG@20. The closer the performance vectors are, the more similar the models are in terms of query by query performance. For the sake of visualization, we reduce the vectors dimension by projecting them to a two-dimensional space, using t-Distributed Stochastic Neighbor Embedding (t-SNE)\footnote{\url{https://lvdmaaten.github.io/tsne/}}.

Figure~\ref{fig:modelproximity} illustrates the proximity of different models in the Robust04 collection. Based on this plot, models with similar input representations (same color) have quite close performance vectors, which means that they perform similarly for same queries. This is not necessarily the case for models with similar architecture (same shape). 
This suggests that the amount and the way that we provide information to the networks are the key factors in the ranking performance. 

We also observe that the \modelone with \feedone is the closest to BM25 which is expected. 
It is also interesting that models with \feedthree are placed far away from other models which shows they perform differently compared to the other input representations.

\mypar{How meaningful are the compositionality weights learned in the \feedthree?}
In this experiment, we focus on the best performing combination, i.e., the \modelthree with \feedthree. To analyze what the network learns, we look into the weights $\mathcal{W}$ (see Section~\ref{sec:feedthree}) learned by the network. Note that the weighting function $\mathcal{W}$ learns a global weight for each vocabulary term. We notice that in both collections there is a strong linear correlation between the learned weights and the inverse document frequency of terms. 
\input{table_res2}
\begin{figure}[!t]%
    \centering
    \begin{subfigure}[t]{0.24\textwidth}
        \centering
        \includegraphics[height=3.3cm]{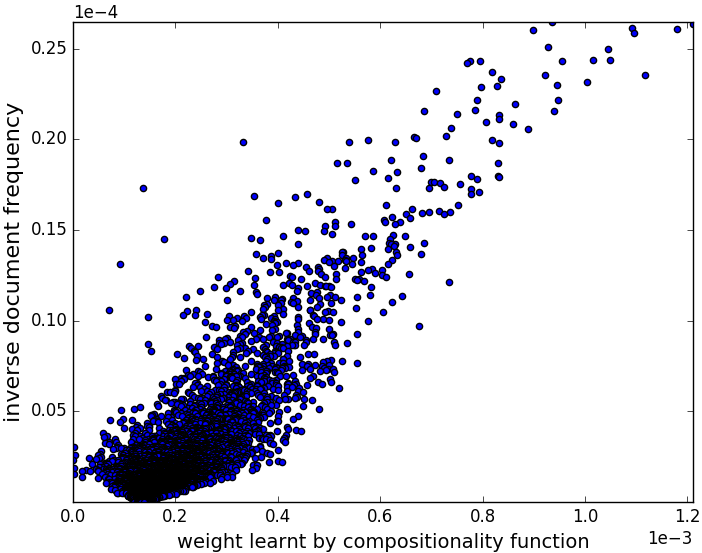}
        \caption{\label{fig:scatter_r}Robust04}{\scriptsize{(Pearson Correlation: 0.8243)}\vspace*{-3ex}}
    \end{subfigure}%
    ~
    \begin{subfigure}[t]{0.24\textwidth}
        \centering
        \includegraphics[height=3.3cm]{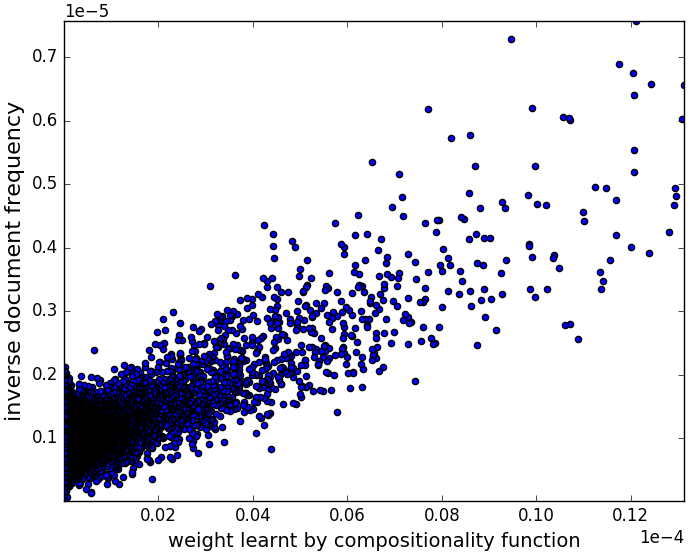}
        \caption{\label{fig:scatter_c}ClueWeb}{\scriptsize{(Pearson Correlation: 0.7014)}\vspace*{-3ex}}
    \end{subfigure}%
    \caption{\label{fig:scatter}Strong linear correlation between weight learned by the compositionality function in the \feedthree and inverse document frequency.}
    \vspace{-10pt}
\end{figure}
Figure~\ref{fig:scatter} illustrates the scatter plots of the learned weight for each vocabulary term and its IDF, in both collections.
This is an interesting observation as we do not provide any global corpus information to the network in training and the network is able to infer such a global information by only observing individual training instances.
% This demonstrates the ability of neural networks to automatically extract meaningful features for the task.

\mypar{How well do other alternatives for the embedding and weighting functions in the \feedthree perform?}\newline 
Considering \feedthree as the input representation, we have examined different alternatives for the embedding function $\mathcal{E}$: (1) employing pre-trained word embeddings learned from an external corpus (we used Google News), (2) employing pre-trained word embeddings learned from the target corpus (using the skip-gram model \cite{Mikolov:2013}), and (3) learning embeddings during the network training as it is explained in Section~\ref{sec:feedthree}. 
Furthermore, for the compositionality function $\odot$, we tried different alternatives: (1) uniform weighting (simple averaging which is a common approach in compositionality function), (2) using IDF as fixed weights instead of learning the weighting function $\mathcal{W}$, and (3) learning weights during the training as described in Section~\ref{sec:feedthree}.

Table~\ref{tbl_res_m3f3_em} presents the performance of all these combinations on both collections. 
We note that learning both embedding and weighting functions leads to the highest performance in both collections. These improvements are statistically significant.
According to the results, regardless of the weighting approach, learning embeddings during training outperforms the models with fixed pre-trained embeddings.
This supports the hypothesis that with the \feedthree the neural networks learn an embedding that is based on the interactions of query and documents that tends to be tuned better to the corresponding ranking task.
Also, regardless of the embedding method, learning weights helps models to get better performance compared to the fixed weightings, with either IDF or uniform weights. 
Although weight learning can significantly affect the performance, it has less impact than learning embeddings.

Note that in the models with pre-trained word embeddings, employing word embeddings trained on the target collection outperforms those trained on the external corpus in the ClueWeb collection; while this is not the case for the Robust04 collection. The reason could be related to the collection size, since the ClueWeb is approximately $100$ times larger than the Robust04.

\begin{figure}[t]
    \centering
    \begin{subfigure}[t]{0.24\textwidth}
        \centering
        \includegraphics[height=3.1cm]{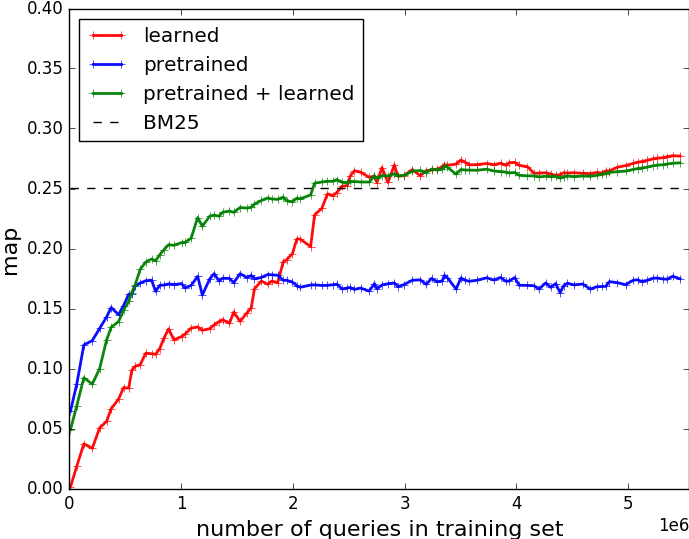}
        \caption{\label{fig:embedding_r}Robust04\vspace*{-3ex}}
    \end{subfigure}%
    ~
    \begin{subfigure}[t]{0.24\textwidth}
        \centering
        \includegraphics[height=3.1cm]{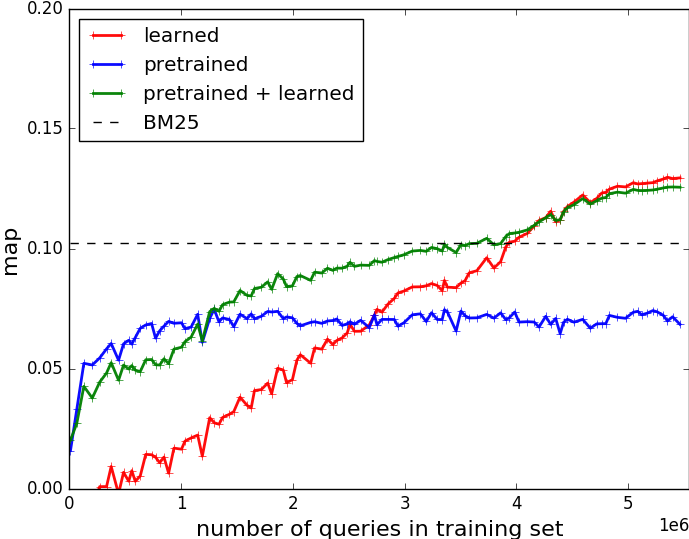}
        \caption{\label{fig:embedding_c}ClueWeb\vspace*{-3ex}}
    \end{subfigure}%
    \caption{\label{fig:embedding}Performance of the \modelthree with learned embedding, pre-trained embedding, and learned embedding with pre-trained embedding as initialization, with respect to different amount of training data.}
    \vspace{-10pt}
\end{figure}
\input{table_res_svm}
\input{table_res_semisup}

In addition to the aforementioned experiments, we have also tried initializing the embedding matrix with a pre-trained word embedding trained on the Google News corpus, instead of random initialization.
Figure~\ref{fig:embedding} presents the learning curve of the models. According to this figure, the model initialized by a pre-trained embedding performs better than random initialization when a limited amount of training data is available. 
When enough training data is fed to the network, initializing with pre-trained embedding and random values converge to the same performance.
An interesting observation here is that in both collections, these two initializations converge when the models exceed the performance of the weak supervision source, which is BM25 in our experiments. 
This suggests that the convergence occurs when accurate representations are learned by the networks, regardless of the initialization.

%In other words, after network sees enough amount of train data to go beyond the supervision signal, it does not matter that with which initialization the model has started to train.

\mypar{Are deep neural networks a good choice for learning to rank with weak supervision?}
To see if there is a real benefit from using a non-linear neural network in different settings, we examined RankSVM~\citep{Joachims:2002} as a strong-performing pair-wise learning to rank method with linear kernel that is fed with different inputs: \feedone, \feedtwo, and \feedthree. Considering that off-the-shelf RankSVM is not able to learn embedding representations during training, for \feedthree, instead of learning embeddings we use a pre-trained embedding matrix trained on Google News and fixed IDF weights. 

The results are reported in Table~\ref{tbl_svm}. As BM25 is not a linear function, RankSVM with linear kernel is not able to completely approximate it. However, surprisingly, for both \feedone and \feedtwo, RankSVM works as well as neural networks (see Table~\ref{tbl_main}). 
Also, compared to the corresponding experiment in Table~\ref{tbl_res_m3f3_em}, the performance of the neural network with an external pre-trained embedding and IDF weighting is not considerably better than RankSVM. 
This shows that having non-linearity in neural networks does not help that much when we do not have representation learning as part of the model.
Note that all of these results are still lower than BM25, which shows that they are not good at learning from weak supervision signals for ranking. 

We have also examined the \modelone with a network with a single linear hidden layer, with the \feedthree, which is equivalent to a linear regression model with the ability of representation learning. 
Comparing the results of this experiment with \mone-\fthree in Table~\ref{tbl_main}, we can see that with a single-linear network we are not able to achieve a performance that is as good as a deep neural network with non-linearity.
This shows that the most important superiority of deep neural networks over other machine learning methods is their ability to learn an effective representation and take all the interactions between query and document(s) into consideration for approximating an effective ranking/scoring function. 
This can be achieved when we have a deep enough network with non-\:linear activations.

%\alexi{Mosi@: we should also include supervised learning experiments on smaller subsets of human judgements, if possible. The idea is that in these scenarios where labeled data is scarce, weak supervision is really useful. Having plots, where we use various percentages of labeled examples and observing how weak supervision is increasingly important as the number of labeled examples becomes smaller.}
\mypar{How useful is learning with weak supervision for supervised ranking?}
In this set of experiments, we investigate whether employing weak supervision as a pre-training step helps to improve the performance of supervised ranking, when a small amount of training data is available. Table~\ref{tbl_semisup} shows the performance of the \modelthree with the \feedthree in three situations: (1) when it is only trained on weakly supervised data (similar to the previous experiments), (2) when it is only trained on supervised data, i.e., relevance judgments, and (3) when the parameters of the network is pre-trained using the weakly supervised data and then fine tuned using relevance judgments.
In all the supervised scenarios, we performed 5-fold cross-\:validation over the queries of each collection and in each step, we used the TREC relevance judgements of the training set as supervised signal. For each query with $m$ relevant documents, we also randomly sampled $m$ non-relevant documents as negative instances. Binary labels are used in the experiments: $1$ for relevant documents and $0$ for non-relevant ones.

The results in Table~\ref{tbl_semisup} suggest that pre-training the network with a weak supervision signal, significantly improves the performance of supervised ranking.
The reason for the poor performance of the supervised model compared to the conventional learning to rank models is that the number of parameters are much larger, hence it needs much more data for training.

In situations when little supervised data is available, it is especially helpful to use unsupervised pre-training which acts as a network pre-conditioning that puts the parameter values in the appropriate range that renders the optimization process more effective for further supervised training~\citep{Rrhan:2010}.

With this experiment, we indicate that the idea of learning from weak supervision signals for neural ranking models, which is presented in this paper, not only enables us to learn neural ranking models when no supervised signal is available, but also has substantial positive effects on the supervised ranking models with limited amount of training data. 

\shrink
\section{Conclusions}
In this paper, we proposed to use traditional IR models such as BM25 as a weak supervision signal in order to generate large amounts of training data to train effective neural ranking models.  
We examine various neural ranking models with different ranking architectures and objectives, and different input representations. 

We used over six million queries to train our models and evaluated them on Robust04 and ClueWeb 09-Category B collections, in an ad-hoc retrieval setting. 
The experiments showed that our best performing model significantly outperforms the BM25 model (our weak supervision signal) by over $13\%$ and $35\%$ MAP improvements in the Robust04 and ClueWeb collections, respectively. 
We also demonstrated that in the case of having a small amount of training data, we can improve the performance of supervised learning by pre-training the network on weakly supervised data.

Based on our results, there are three key ingredients in neural ranking models that lead to good performance with weak supervision:
The first is the proper input representation. Providing the network with raw data and letting the network to learn the features that matter, gives the network a chance of learning how to ignore imperfection in the training data.
The second ingredient is to target the right goal and define a proper objective function. In the case of having weakly annotated training data, by targeting some explicit labels from the data, we may end up with a model that learned to express the data very well, but is incapable of going beyond it. 
This is especially the case with deep neural networks where there are many parameters and it is easy to learn a model that overfits the data.
The third ingredient is providing the network with a considerable amount of training examples. 
As an example, during the experiments we noticed that using the \feedthree, the network needs a lot of examples to learn embeddings that are more effective for retrieval compared to pre-trained embeddings. 
Thanks to weak supervision, we can generate as much training data as we need with almost no cost.

%Diversity: in pair-wise models, they need to get both soft and hard negative examples during training to keep the model learning at the edge of its ability. 

Several future directions can be pursued. 
An immediate task would be to study the performance of more expressive neural network architectures e.g., CNNs and LSTMs, with weak supervision setup.  
Other experiment is to leverage multiple weak supervision signals from different sources. For example, we have other unsupervised ranking signals such as query likelihood and PageRank and taking them into consideration might benefit the learning process. 
Other future work would be to investigate the boosting mechanism for the method we have in this paper. In other words, it would be interesting to study if it is possible to use the trained model on weakly supervised data to annotate data with more quality from original source of annotation and leverage the new data to train a better model. 
Finally, we can apply this idea to other information retrieval tasks, such as query/document classification and clustering.

% \mypar{Acknowledgments}
% This research is funded in part by Netherlands Organization for Scientific Research through the \textsl{Exploratory Political Search} project (ExPoSe, NWO CI \# 314.99.108), and by the Digging into Data Challenge through the \textsl{Digging Into Linked Parliamentary Data} project (DiLiPaD, NWO Digging into Data \# 600.006.014).

\shrink
\begin{acks}
%The first author would like to thank Sascha Rothe for his valuable comments and helpful suggestions.
This research was supported in part by Netherlands Organization for Scientific Research through the \textsl{Exploratory Political Search} project (ExPoSe, NWO CI \# 314.99.108), by the Digging into Data Challenge through the \textsl{Digging Into Linked Parliamentary Data} project (DiLiPaD, NWO Digging into Data \# 600.006.014), and by the Center for Intelligent Information Retrieval. Any opinions, findings and conclusions or recommendations expressed in this material are those of the authors and do not necessarily reflect those of the sponsors.
\end{acks}

\sshrink
\bibliographystyle{ACM-Reference-Format}
\bibliography{ref} 

\end{document}

%% file: table_data.tex
% !TEX root = main.tex
\begin{table}
\centering
\caption{Collections statistics.}
\vspace{-10pt}
\begin{tabularx}{\linewidth}{Xcccc} 
\toprule
\bf Collection & \bf Genre & \bf Queries & \bf \# docs & \bf length  \\ %\specialcell{avg doc\\length}  \\ 
\midrule
%Robust & news articles & \specialcell{301-450 \\ 601-700} & 528k & 254  \\ %\hline
\bf Robust04 & news  & 301-450,601-700 & 528k & 254  \\ %\hline
\bf ClueWeb & webpages & 1-200 & 50m & 1,506  \\ 
\bottomrule
% GOV2 & \specialcell{2004 crawl of .gov domains} & \specialcell{TREC 2004-2006 Terabyte Track,\\topics 701-850} & 25,205k  & 648 & 26,917  \\ \hline
\end{tabularx}
\label{tab:data}
\vspace{-15pt}
\end{table}

%% file: table_res.tex
% !TEX root = main.tex
\newcommand{\ps}{$^\blacktriangleup$}
\newcommand{\ns}{$^\smalltriangledown$}
\newcommand{\fs}{$^{~}$}

\begin{table*}[tbp]
\centering
\caption{\label{tbl_main}Performance of the different models on different datasets. \ps or \ns indicates that the improvements or degradations with respect to BM25 are statistically significant, at the 0.05 level using the paired two-tailed t-test.}
\vspace{-5pt}
%\begin{adjustbox}{max width=\columnwidth}
%\begin{tabular}{@{}l@{~~~}c@{~~}c@{~~}c@{~~~}c@{~~}c@{~~}c@{}}
\begin{adjustbox}{max width=0.9\textwidth}
\begin{tabular}{l c c c c c c}
\toprule
\multirow{2}{*}{\textbf{Method}} &
%\multicolumn{1}{l}{\textbf{Method}} & 
\multicolumn{3}{c}{\textbf{Robust04}} & \multicolumn{3}{c}{\textbf{ClueWeb}}
\\ \cmidrule(lr){2-4} \cmidrule(lr){5-7}
& \textit{MAP} & \textit{P@20} & \textit{nDCG@20}  & \textit{MAP} & \textit{P@20} & \textit{nDCG@20}
\\ \midrule
\textbf{BM25} 
& 0.2503\fs & 0.3569\fs & 0.4102\fs  
& 0.1021\fs & 0.2418\fs & 0.2070\fs
\\ \midrule
\textbf{\mone + \fone} 
& 0.1961\ns & 0.2787\ns & 0.3260\ns 
& 0.0689\ns & 0.1518\ns & 0.1430\ns
\\ 
\textbf{\mone + \ftwo} 
& 0.2141\ns & 0.3180\ns & 0.3604\ns 
& 0.0701\ns & 0.1889\ns & 0.1495\ns
\\ 
\textbf{\mone + \fthree} 
& 0.2423\ns & 0.3501\fs & 0.3999\fs 
& 0.1002\fs & 0.2513\fs & 0.2130\fs
\\ \midrule
\textbf{\mtwo + \fone} 
& 0.1940\ns & 0.2830\ns & 0.3317\ns 
& 0.0622\ns & 0.1516\ns & 0.1383\ns
\\ 
\textbf{\mtwo + \ftwo} 
& 0.2213\ns & 0.3216\ns & 0.3628\ns 
& 0.0776\ns & 0.1989\ns & 0.1816\ns
\\ 
\textbf{\mtwo + \fthree} 
& \textbf{0.2811}\ps & \textbf{0.3773}\ps & \textbf{0.4302}\ps 
& \textbf{0.1306}\ps & \textbf{0.2839}\ps & \textbf{0.2216}\ps
\\ \midrule
\textbf{\mthree + \fone} 
& 0.2192\ns & 0.2966\ns & 0.3278\ns 
& 0.0702\ns & 0.1711\ns & 0.1506\ns
\\ 
\textbf{\mthree + \ftwo} 
& 0.2246\ns & 0.3250\ns & 0.3763\ns 
& 0.0894\ns & 0.2109\ns & 0.1916\fs
\\ 
\textbf{\mthree + \fthree} 
& \textbf{0.2837}\ps & \textbf{0.3802}\ps & \textbf{0.4389}\ps 
& \textbf{0.1387}\ps & \textbf{0.2967}\ps & \textbf{0.2330}\ps
\\ \bottomrule
\end{tabular}
\end{adjustbox}
\vspace{-5pt}
\end{table*}

%% file: plot_loss_step.tex
% !TEX root = main.tex
\begin{figure*}[t]
    \centering
    \begin{subfigure}[t]{0.3\textwidth}
        \centering
        \includegraphics[height=3.5cm]{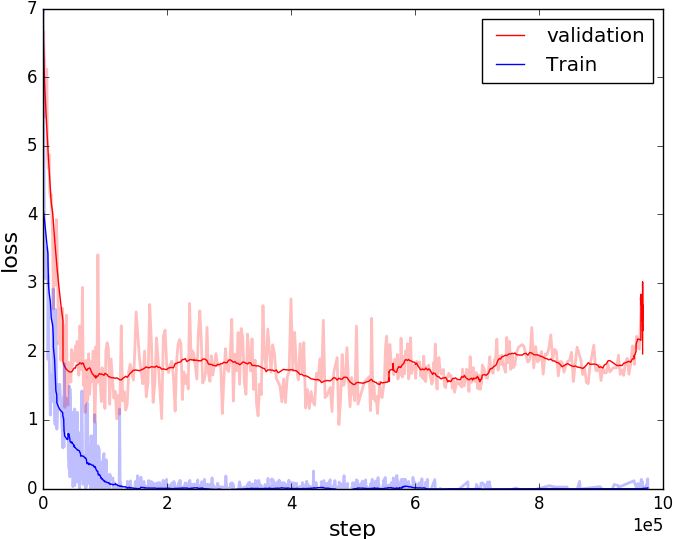}
        \caption{\label{fig:m1f1}\mone-\fone}
    \end{subfigure}%
    ~
    \begin{subfigure}[t]{0.3\textwidth}
        \centering
        \includegraphics[height=3.5cm]{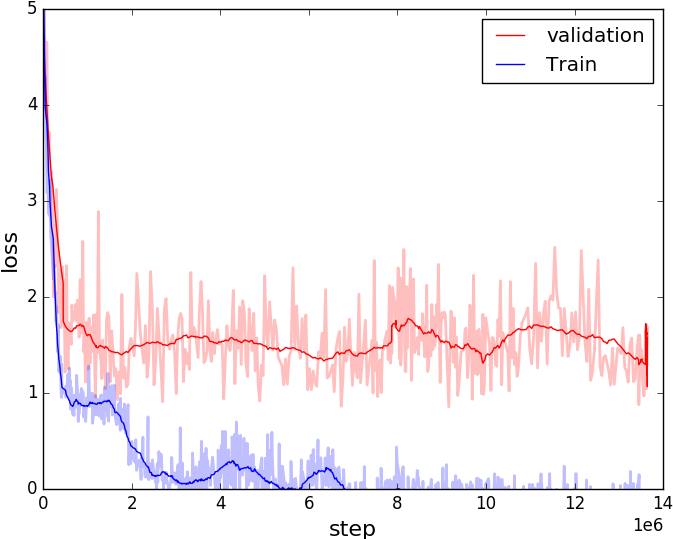}
        \caption{\label{fig:m1f2}\mone-\ftwo}
    \end{subfigure}%
    ~
    \begin{subfigure}[t]{0.3\textwidth}
        \centering
        \includegraphics[height=3.5cm]{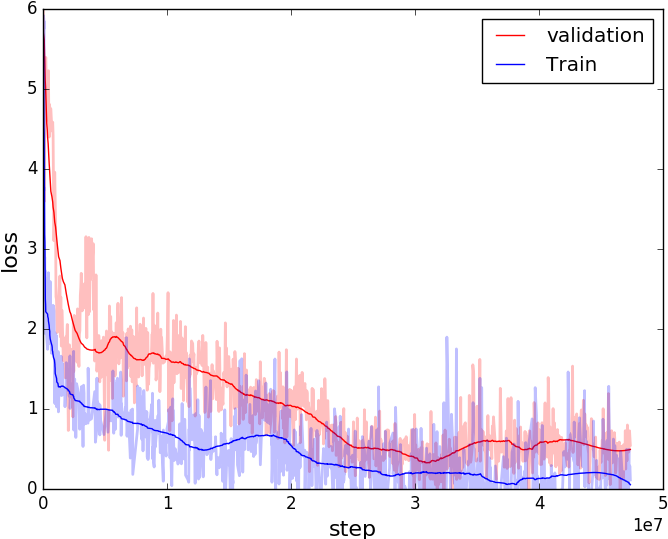}
        \caption{\label{fig:m1f3}\mone-\fthree}
    \end{subfigure}%
    \\
    \begin{subfigure}[t]{0.3\textwidth}
        \centering
        \includegraphics[height=3.5cm]{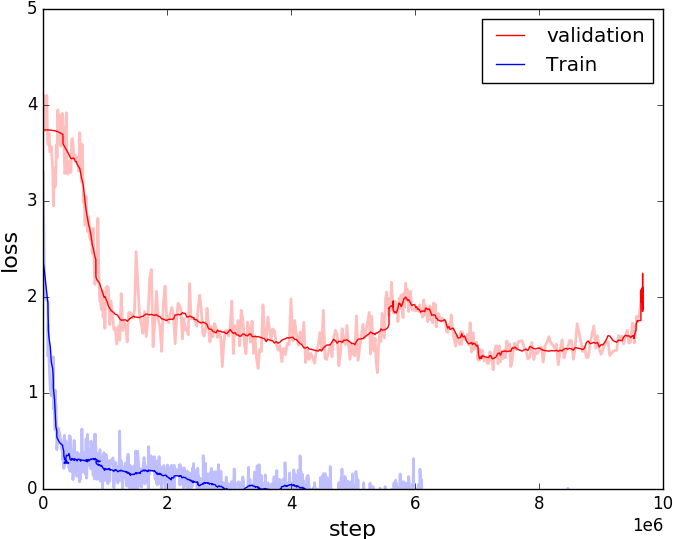}
        \caption{\label{fig:m2f1}\mtwo-\fone}
    \end{subfigure}%
    ~
    \begin{subfigure}[t]{0.3\textwidth}
        \centering
        \includegraphics[height=3.5cm]{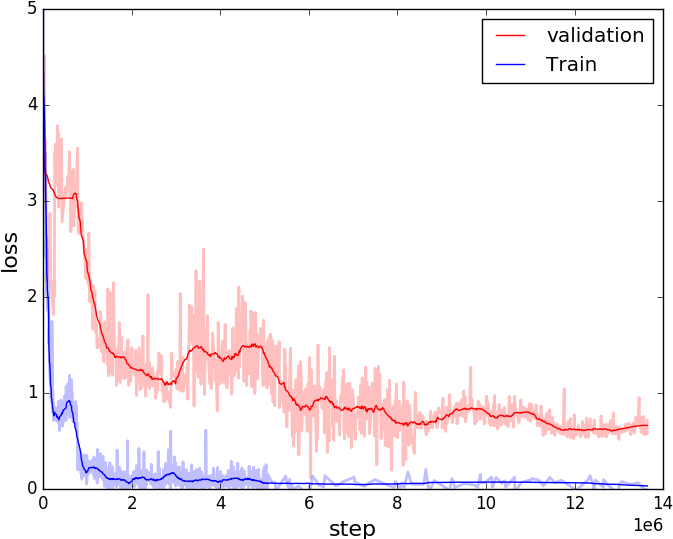}
        \caption{\label{fig:m2f2}\mtwo-\ftwo}
    \end{subfigure}%
    ~
    \begin{subfigure}[t]{0.3\textwidth}
        \centering
        \includegraphics[height=3.5cm]{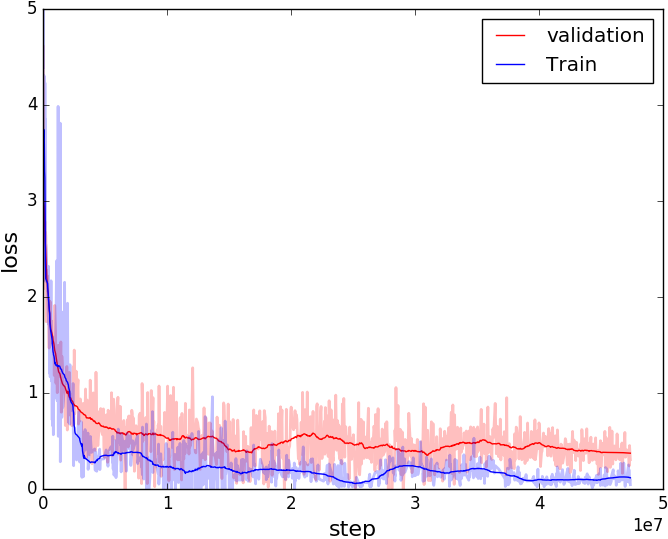}
        \caption{\label{fig:m2f3}\mtwo-\fthree}
    \end{subfigure}%
        \\
    \begin{subfigure}[t]{0.3\textwidth}
        \centering
        \includegraphics[height=3.5cm]{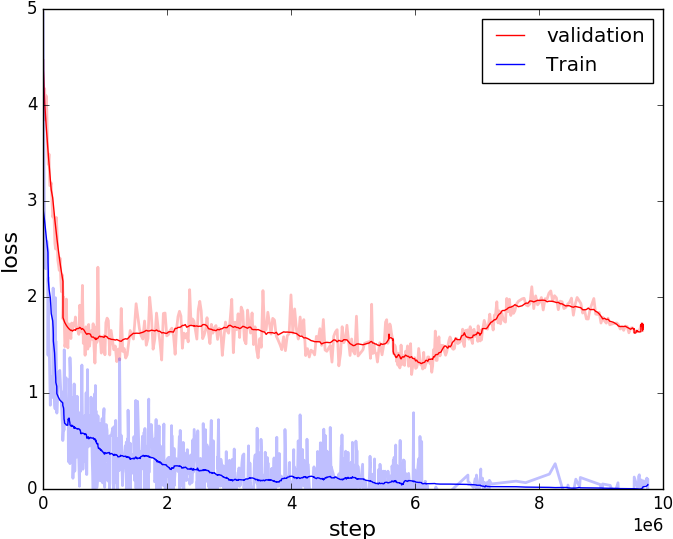}
        \caption{\label{fig:m3f1}\mthree-\fone}
    \end{subfigure}%
    ~
    \begin{subfigure}[t]{0.3\textwidth}
        \centering
        \includegraphics[height=3.5cm]{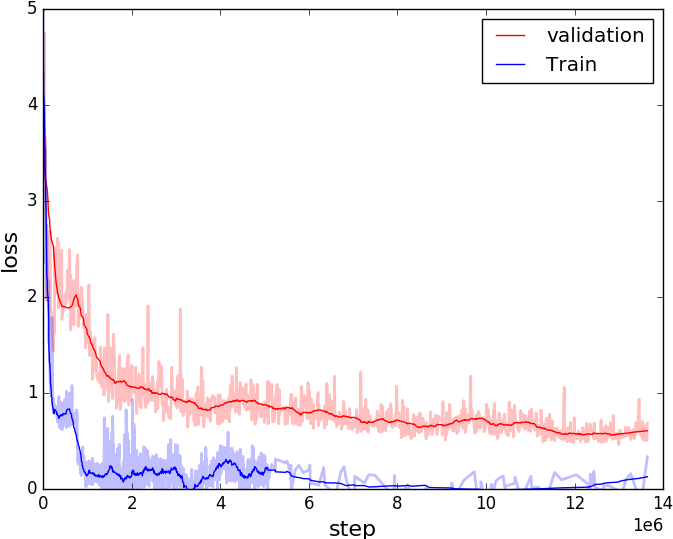}
        \caption{\label{fig:m3f2}\mthree-\ftwo}
    \end{subfigure}%
    ~
    \begin{subfigure}[t]{0.3\textwidth}
        \centering
        \includegraphics[height=3.5cm]{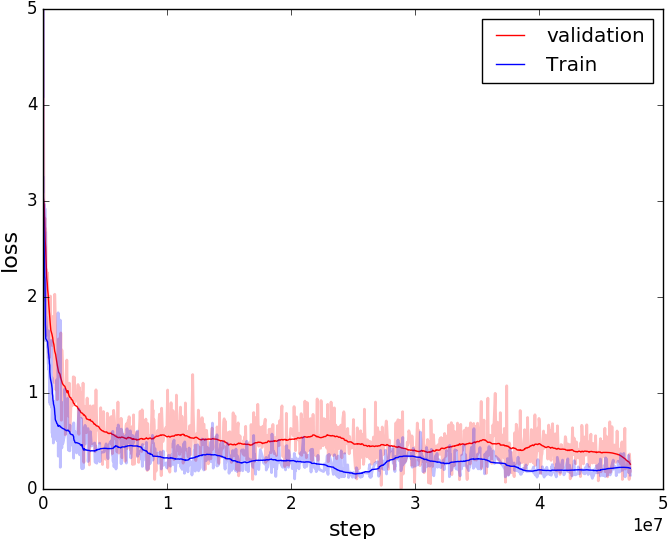}
        \caption{\label{fig:m3f3}\mthree-\fthree}
    \end{subfigure}%
    \vspace{-5pt}
    \caption{Training and validation loss curves for all combinations of different ranking architectures and feeding paradigms.}
    \label{fig:step-loss}
    \vspace{-5pt}
\end{figure*}

%% file: models_proximity.tex
% !TEX root = main.tex
\begin{figure}[!t]
\pgfplotsset{ticks=none}
\centering
\begin{tikzpicture}
\begin{axis}[
        width= 9cm, 
        height=5.5cm,
        grid = both,
        nodes near coords, %={(\coordindex)},
        scatter/classes={
        	BM25={mark=otimes*, mark size=3pt, yellow, draw = black},%
        	m1f1={mark=halfcircle*, mark size=3pt, blue, },%
        	m1f2={mark=halfcircle*, mark size=3pt, red},%
        	m1f3={mark=halfcircle*, mark size=3pt, black},
        	m2f1={mark=halfsquare*, mark size=3pt, blue},%
        	m2f2={mark=halfsquare*, mark size=3pt, red},%
        	m2f3={mark=halfsquare*, mark size=3pt, black},
        	m3f1={mark=pentagon*, mark size=3pt, blue},%
        	m3f2={mark=pentagon*, mark size=3pt, red},%
        	m3f3={mark=pentagon*, mark size=3pt, black}
    	},
    	legend style={
            % legend image post style={xscale=0.75},
            % inner sep=0pt,
            at={(1.2,0.85)},
            font=\fontsize{6}{7}\selectfont,
        },
    	]
	\addplot[scatter,only marks,
		scatter src=explicit symbolic]
		coordinates {
			(0.403,0.801) [BM25]
			(0.341,0.749) [m1f1]
			(0.232,0.240) [m1f2]
			(0.570,0.412) [m1f3]
			(0.091,0.718) [m2f1]
			(0.265,0.369) [m2f2]
			(0.808,0.101) [m2f3]
			(0.246,0.601) [m3f1]
			(0.371,0.600) [m3f2]
			(0.730,0.305) [m3f3]
		};
	\node [right] at (0.413,0.801){\footnotesize{BM25}};
	\node [left] at (0.381,0.800){\footnotesize{\mone + \fone}};
	\node [below] at (0.232,0.230){\footnotesize{\mone + \ftwo}};
	\node [above] at (0.570,0.422){\footnotesize{\mone + \fthree}};
	\node [below] at (0.121,0.708){\footnotesize{\mtwo + \fone}};
	\node [above] at (0.265,0.379){\footnotesize{\mtwo + \ftwo}};
	\node [above] at (0.758,0.121){\footnotesize{\mtwo + \fthree}};
	\node [below] at (0.226,0.591){\footnotesize{\mthree + \fone}};
	\node [right] at (0.381,0.600){\footnotesize{\mthree + \ftwo}};
	\node [above] at (0.730,0.315){\footnotesize{\mthree + \fthree}};
%  	\legend{BM25,m1f1,m1f2,m1f3,m2f1,m2f2,m2f3,m3f1,m3f2,m3f3}
\end{axis}
\end{tikzpicture}
\vspace{-5pt}
\caption{Proximity of different models in terms of query-by-query performance.
 \label{fig:modelproximity}}
 \vspace{-15pt}
 \end{figure}

%% file: table_res2.tex
% !TEX root = main.tex
\begin{table*}[tbp]
\centering
\caption{\label{tbl_res_m3f3_em}Performance of the \modelthree with variants of the \feedthree on different datasets. \ps indicates that the improvements over all other models are statistically significant, at the 0.05 level using the paired two-tailed t-test, with Bonferroni correction.}
\vspace{-5pt}
%\begin{adjustbox}{max width=\columnwidth}
%\begin{tabular}{@{}l@{~~~}c@{~~}c@{~~}c@{~~~}c@{~~}c@{~~}c@{}}
\begin{adjustbox}{max width=0.9\textwidth}
\begin{tabular}{l c c c c c c}
\toprule
\multirow{2}{*}{\textbf{Embedding type}} &
%\multicolumn{1}{l}{\textbf{Method}} & 
\multicolumn{3}{c}{\textbf{Robust04}} & \multicolumn{3}{c}{\textbf{ClueWeb}}
\\ \cmidrule(lr){2-4} \cmidrule(lr){5-7}
& \textit{MAP} & \textit{P@20} & \textit{nDCG@20}  & \textit{MAP} & \textit{P@20} & \textit{nDCG@20}
\\ \midrule
\textbf{Pretrained (external) + Uniform weighting} 
& 0.1656 & 0.2543 & 0.3017 
& 0.0612 & 0.1300 & 0.1401
\\ 
\textbf{Pretrained (external) + IDF weighting} 
& 0.1711 & 0.2755 & 0.3104 
& 0.0712 & 0.1346 & 0.1469
\\ 
\textbf{Pretrained (external) + Weight learning} 
& 0.1880 & 0.2890 & 0.3413 
& 0.0756 & 0.1344 & 0.1583
\\ 
\textbf{Pretrained (target) + Uniform weighting} 
& 0.1217 & 0.2009 & 0.2791 
& 0.0679 & 0.1331 & 0.1587
\\ 
\textbf{Pretrained (target) + IDF weighting} 
& 0.1402 & 0.2230 & 0.2876 
& 0.0779 & 0.1674 & 0.1540
\\ 
\textbf{Pretrained (target) + Weight learning} 
& 0.1477 & 0.2266 & 0.2804 
& 0.0816 & 0.1729 & 0.1608
\\
\textbf{Learned + Uniform weighting} 
& 0.2612 & 0.3602 & 0.4180 
& 0.0912 & 0.2216 & 0.1841
\\
\textbf{Learned + IDF weighting} 
& 0.2676 & 0.3619 & 0.4200 
& 0.1032 & 0.2419 & 0.1922
\\ 
\textbf{Learned + Weight learning} 
& \textbf{0.2837}\ps & \textbf{0.3802}\ps & \textbf{0.4389}\ps
& \textbf{0.1387}\ps & \textbf{0.2967}\ps & \textbf{0.2330}\ps
\\ \bottomrule
\end{tabular}
\end{adjustbox}
\vspace{-5pt}
\end{table*}

%% file: table_res_svm.tex
% !TEX root = main.tex
\begin{table*}[tbp]
\centering
\caption{\label{tbl_svm}Performance of the linear RankSVM with different features.}
\vspace{-10pt}
%\begin{adjustbox}{max width=\columnwidth}
%\begin{tabular}{@{}l@{~~~}c@{~~}c@{~~}c@{~~~}c@{~~}c@{~~}c@{}}
\begin{adjustbox}{max width=0.8\textwidth}
\begin{tabular}{l c c c c c c}
\toprule
\multirow{2}{*}{\textbf{Method}} &
%\multicolumn{1}{l}{\textbf{Method}} & 
\multicolumn{3}{c}{\textbf{Robust04}} & \multicolumn{3}{c}{\textbf{ClueWeb}}
\\ \cmidrule(lr){2-4} \cmidrule(lr){5-7}
& \textit{MAP} & \textit{P@20} & \textit{nDCG@20}  & \textit{MAP} & \textit{P@20} & \textit{nDCG@20}
\\ \midrule
\textbf{RankSVM + \fone} 
& 0.1983\fs & 0.2841\fs & 0.3375\fs 
& 0.0761\fs & 0.1840\fs & 0.1637\fs
\\ 
\textbf{RankSVM + \ftwo} 
& 0.2307\fs & 0.3260\fs & 0.3794\fs 
& 0.0862\fs & 0.2170\fs & 0.1939\fs
\\ 
\textbf{RankSVM + (Pretrained (external) + IDF weighting)} 
& 0.1539\fs & 0.2121\fs & 0.1852\fs 
& 0.0633\fs & 0.1572\fs & 0.1494\fs 
\\ \midrule
\textbf{\mone (one layer with no nonlinearity) + \fthree} 
& 0.2103\fs & 0.3986\fs & 0.3160\fs 
& 0.0645\fs & 0.1421\fs & 0.1322\fs
\\ \bottomrule
\end{tabular}
\end{adjustbox}
\vspace{-5pt}
\end{table*}

%% file: table_res_semisup.tex
% !TEX root = main.tex
\begin{table*}[tbp]
\centering
\caption{Performance of the \modelthree with \feedthree in fully supervised setting, weak supervised setting, and weak supervised plus supervision as fine tuning. \ps indicates that the improvements over all other models are statistically significant, at the 0.05 level using the paired two-tailed t-test, with Bonferroni correction.}
\label{tbl_semisup}
\vspace{-10pt}
%\begin{adjustbox}{max width=\columnwidth}
%\begin{tabular}{@{}l@{~~~}c@{~~}c@{~~}c@{~~~}c@{~~}c@{~~}c@{}}
\begin{adjustbox}{max width=0.8\textwidth}
\begin{tabular}{l c c c c c c}
\toprule
\multirow{2}{*}{\textbf{Method}} &
%\multicolumn{1}{l}{\textbf{Method}} & 
\multicolumn{3}{c}{\textbf{Robust04}} & \multicolumn{3}{c}{\textbf{ClueWeb}}
\\ \cmidrule(lr){2-4} \cmidrule(lr){5-7}
& \textit{MAP} & \textit{P@20} & \textit{nDCG@20}  & \textit{MAP} & \textit{P@20} & \textit{nDCG@20}
\\ \midrule
\textbf{Weakly supervised} 
& 0.2837 \fs & 0.3802\fs & 0.4389\fs  
& 0.1387 \fs & 0.2967\fs & 0.2330\fs
\\
\textbf{Fully supervised} 
& 0.1790 \fs & 0.2863\fs & 0.3402\fs  
& 0.0680 \fs & 0.1425\fs & 0.1652\fs
\\
\textbf{Weakly supervised + Fully supervised} 
& \textbf{0.2912}\ps & \textbf{0.4126}\ps & \textbf{0.4509}\ps 
& \textbf{0.1520}\ps & \textbf{0.3077}\ps & \textbf{0.2461}\ps
\\ \bottomrule
\end{tabular}
\end{adjustbox}
\vspace{-5pt}
\end{table*}